\documentclass[aps,prl,10pt,superscriptaddress,showpacs,amsmath,twocolumn,floatfix]{revtex4}
\usepackage{amsmath}
\usepackage{amsfonts}
\usepackage{graphicx}
\usepackage{color}

\newcommand{\real}{\operatorname{Re}}

\newcommand{\parti}[2]{\frac{\partial #1}{\partial #2}}

\newcommand{\intall}{\int_{-\infty}^{\infty}}

\newcommand{\abs}[1]{\left|#1\right|}
\newcommand{\bk}[1]{\left(#1\right)}

\newcommand{\trace}{\operatorname{tr}}

\begin{document}

\title{Fundamental Quantum Limit to Waveform Estimation}

\author{Mankei Tsang}
\email{mankei@unm.edu} \affiliation{Center for Quantum Information
and Control, University of New Mexico, MSC07--4220, Albuquerque, New
Mexico 87131-0001, USA}

\author{Howard M.~Wiseman}
\affiliation{Centre for Quantum Computation and Communication
Technology (Australian Research Council),
\\ Centre for Quantum Dynamics, Griffith University,
Brisbane, Queensland 4111, Australia}

\author{Carlton M.~Caves}
\affiliation{Center for Quantum Information and Control, University
of New Mexico, MSC07--4220, Albuquerque, New Mexico 87131-0001, USA}

\date{\today}

\begin{abstract}
  We derive a quantum Cram\'er-Rao bound (QCRB) on the error of
  estimating a time-changing signal.  The QCRB provides a fundamental
  limit to the performance of general quantum sensors, such as
  gravitational-wave detectors, force sensors, and atomic
  magnetometers.  We apply the QCRB to the problem of force
  estimation via continuous monitoring of the position of a harmonic
  oscillator, in which case the QCRB takes the form of a spectral
  uncertainty principle.  The bound on the force-estimation error can
  be achieved by implementing quantum noise cancellation in the
  experimental setup and applying smoothing to the observations.
\end{abstract}

\maketitle

The accuracy of any sensor is limited by noise.  To quantify the
potential performance of a sensor, it is often useful to compute a
lower bound to the error in the estimation of the signal of interest.
One of the most widely used bounds is the \emph{Cram\'er-Rao bound\/}
(CRB), which limits the mean-square error in parameter
estimation~\cite{vantrees}.

The development of quantum technology highlights the question of how
quantum mechanics impacts the performance of sensors. Helstrom
formulated a \emph{quantum Cram\'er-Rao bound\/}
(QCRB)~\cite{helstrom}, which stipulates that the minimum estimation
error is inversely proportional to a property of the sensor known as
the quantum Fisher information.  The QCRB is central to quantum
sensor design in the burgeoning field of quantum
metrology~\cite{wiseman,glm} for several reasons.  It allows one to
determine whether the fundamental sensitivity of a sensor design
meets the requirements of an application, provides a criterion
against which the optimality of quantum sensing schemes can be
tested, and motivates improvements of schemes that are suboptimal.
For sensors near the fundamental limit, the QCRB can also be used to
quantify the trade-off between sensing accuracy and physical
resources of the sensor, so that efficient ways of improving
sensitivity can be identified.

Most prior work on the QCRB considered estimation of one or a few
fixed parameters.  Yet, in most sensing applications, such as force
sensing and magnetometry, the signal of interest is changing in time.
This time-changing signal, which we call a \textit{waveform}, is
coupled continuously to the sensor, and continuous measurements on the
sensor are used to extract information about the
waveform~\cite{braginsky,berry,smooth}.  Here we derive the QCRB for
waveform estimation---the first such derivation to our
knowledge---allowing for any quantum measurement protocol, including
sequential, discrete or continuous measurements.

Previous work on the QCRB generally did not take into account prior
information, but for the task of estimating a waveform, which often
depends on an infinite number of unknown parameters, parameter estimation
techniques no longer suffice and prior information is required to make
the problem well defined \cite{vantrees}.  The prior information
might, for example, restrict the signal to a finite bandwidth, making
integrals over frequency finite that otherwise would diverge.  Thus a
crucial feature of our QCRB is the inclusion of prior waveform
information.

Our result provides a rigorous criterion against which the optimality
of design, control, and estimation strategies for quantum sensors,
such as gravitational-wave detectors, force sensors, and atomic
magnetometers, can be tested. As an example, we calculate the QCRB on
the error of force estimation via continuous position measurements of
a harmonic oscillator, in which case the bound takes the form of a
\emph{spectral uncertainty principle}.  We show that the bound can be
achieved by implementing \emph{quantum noise cancellation\/} (QNC) to
remove the backaction noise from the observations~\cite{qnc} and
applying the estimation technique of \emph{quantum
smoothing\/}~\cite{smooth} to the observations.  This proves the
optimality of such control and estimation techniques for force
sensing and establishes our QCRB as the fundamental limit to force
sensing.

Let $x(t)$ denote the classical waveform to be estimated.  For
simplicity, we assume $x(t)$ to be a scalar function; generalization
to multiple processes is straightforward.  We discretize time as $t_j
= t_0+j\delta t$, $j=0,1,\dots,J$, and assume that $\delta t$ is
small enough that we can treat $x(t)$ as piecewise-constant, i.e.,
$x(t) = x_j$ for $t_{j}\le t < t_{j+1}$. The prior probability
density $P[x]$ for the vector $x\equiv(x_{J-1},\dots,x_0)^T$
characterizes what is known or assumed about the waveform prior to
the measurements.  For a vector of observations
$y\equiv(y_{N-1},\dots,y_1,y_0)^T$ made any time during the interval
$t_0< t \le t_J$, we define a conditional probability density
$P[y|x]$.  The joint probability density is $P[y,x] = P[y|x]P[x]$.
Finally, we define the estimate of $x_j$ as $\tilde x_j[y]$ and the
estimate bias, given signal $x$, as $\int\!Dy\,(\tilde
x_j-x_j)P[y|x]\equiv b_j[x]$, where $Dy \equiv \prod_{n=0}^{N-1}
dy_n$.

Multiplying both sides of $b_j[x]$ by $P[x]$, differentiating with
respect to $x_k$, and then integrating over all $x$ using
$Dx\equiv\prod_{j=0}^{J-1}dx_j$, we obtain
\begin{align}
-\delta_{jk}+
\int\! Dx&\,Dy\,(\tilde x_j - x_j)\parti{P[y,x]}{x_k}\nonumber\\
&=\int\! Dx \parti{}{x_k}\bigl(b_j[x]P[x]\bigr)=0\;,
\label{step}
\end{align}
where the final equality assumes $b_j[x]P[x]|_{x_k=\pm\infty} = 0$.
This assumption, also used in the proof of the classical CRB
\cite{vantrees}, is satisfied as long as the prior density approaches
zero at the infinite endpoints (as it must for any probability
density) and the bias there is not infinite.

\begin{figure}[htbp]
\centerline{\includegraphics[width=0.48\textwidth]{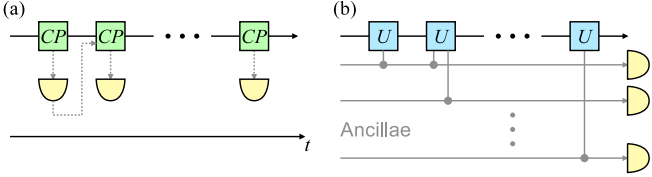}}
\caption{Any quantum dynamics and sequential measurements described
  by completely positive (CP) maps, including feedback
  based on measurement outcomes, as illustrated in~(a), can be
  reproduced by unitary evolution of an enlarged system that includes
  appropriate ancillae, coherent controlled unitaries, and deferred
  measurements of the ancillae, as shown in~(b).}
\label{naimark}
\end{figure}

Quantum mechanics enters this description, which till now is
classical, by determining the conditional probability of the
observations.  Given a quantum system, we can describe any
measurement protocol, including sequential measurements and excess
decoherence, during the interval $t_0\le t < t_J$ by introducing
appropriate ancillae, in accord with the Kraus representation
theorem~\cite{wiseman,kraus,nielsen}. This also accounts for any
feedback during the interval, based on the measurement outcomes,
because the principle of deferred measurement~\cite{nielsen} allows
one to put off the measurements on the ancillae till time $t_J$;
measurement-based feedback is replaced by controlled unitaries prior
to the measurements, as schematically shown in Fig.~\ref{naimark}.

In this approach, the overall system dynamics is described by unitary
evolution of the enlarged system; the conditional probability of
observations is given by $P[y|x]=\trace\bigl(E[y]\rho_x\bigr)$, where
$\rho_x$ is the density operator of the enlarged system at time
$t_J$, conditioned upon $x$, and $E[y]$ is the
positive-operator-valued measure (POVM) that describes the (deferred)
measurements up to time $t_J$.  We denote expectation values with
respect to $\rho_x$ by angle brackets subscripted by~$x$, so that
$\langle E[y]\rangle_x\equiv \trace\bigl(E[y]\rho_x\bigr)$.
Continuous measurements can be modeled as the limit of a sequence of
infinitesimally weak measurements~\cite{wiseman}.


We now follow a procedure similar to the one used by Helstrom
\cite{helstrom} to derive the \hbox{QCRB}.  We introduce an operator
$Q_k$ that satisfies $\partial\rho_x/\partial
x_k=(Q_k\rho_x+\rho_xQ_k^\dagger)/2$.
Unlike Helstrom, we do not require $Q_k$ to be Hermitian. Note that
the vanishing trace of $\partial\rho_x/\partial x_k$ in the
definition of $Q_k$ implies that $\real\langle Q_k\rangle_x=0$.

It is convenient to incorporate the prior information by working in
terms of a density operator $\rho[x] \equiv\rho_x P[x]$ in a hybrid
quantum-classical space and introducing an operator
$L_k=Q_k+\partial\ln P[x]/\partial x_k$, which satisfies
$\partial\rho[x]/\partial
x_k=(L_k[x]\rho[x]+\rho[x]L_k^\dagger[x])/2$.  In terms of $L_k$,
Eq.~(\ref{step}) takes the form that we use to derive the QCRB:
\begin{align}
\delta_{jk}=
\real\int\! Dx\,Dy\,(\tilde x_j-x_j)\trace\bigl(E[y]L_k[x]\rho[x]\bigr)\;.
\label{step2}
\end{align}
Multiplying Eq.~(\ref{step2}) by $u_jv_k$, where $u_j$ and $v_k$ are
the components of arbitrary real column vectors $u$ and $v$, and then
summing over all $j$ and $k$, we obtain
\begin{align}
v^Tu=\sum_j u_j v_j=\real\int\! Dx\,Dy\,\trace\!\bk{A^\dagger B}\;,
\label{vu}
\end{align}
where $A^\dagger \equiv \sum_k v_k \sqrt{E[y]}L_k\sqrt{\rho[x]}$, $B
\equiv \sum_j u_j (\tilde x_j-x_j)\sqrt{\rho[x]}\sqrt{E[y]}$, and $T$
denotes transposition. It follows from Eq.~(\ref{vu}) that
\begin{align}
(v^Tu)^2&\le
\abs{\int\! Dx\,Dy\,\trace\!\bk{A^\dagger B}}^2\nonumber\\
&\le
\int\! Dx\,Dy\,\trace(A^\dagger A)
\int\! Dx\,Dy\,\trace(B^\dagger B)\;,
\label{step3}
\end{align}
where the second inequality is the Schwarz inequality.

The second integral in Eq.~(\ref{step3}) is $\int\!
Dx\,Dy\,\trace(B^\dagger B)=u^T\Sigma\,u$, where
\begin{align}
\Sigma_{jk}\equiv\int\! Dx\,Dy\,P[x,y](\tilde x_j-x_j)(\tilde x_k-x_k)
\label{Sigmajk}
\end{align}
is the estimation-error covariance matrix.  The first integral in
Eq.~(\ref{step3}) is, using the completeness of the POVM, $\int\!
Dx\,Dy\,\trace(A^\dagger A)=v^TFv$, where $F$ is a (real, symmetric)
Fisher-information matrix,
\begin{align}
F_{jk}\equiv\frac{1}{2}
\int\! Dx\,P[x]\trace\Bigl((L_j^\dagger L_k+L_k^\dagger L_j)\rho_x\Bigr)\;.
\label{fisher}
\end{align}
Since $\real \langle Q_k\rangle_x=0$, $F$ separates neatly into a quantum
and a classical, prior-information component, i.e.,
$F=F^{(Q)}+F^{(C)}$, where
\begin{align}
F_{jk}^{(Q)}&=\frac{1}{2}
\int\! Dx\,P[x]\trace\Bigl((Q_j^\dagger Q_k+Q_k^\dagger Q_j)\rho_x\Bigr)
\label{FQ}\\
F_{jk}^{(C)}&=
\int\! Dx\,
P[x]\parti{\ln P[x]}{x_j}\parti{\ln P[x]}{x_k}\;.
\end{align}

When these results are substituted into Eq.~(\ref{step3}), we find
that $(v^T F v)(u^T \Sigma u)\ge(v^T u)(u^T v)$.  Setting $v = F^{-1}
u$ implies that $u^T(\Sigma-F^{-1})u\ge0$ for arbitrary real vectors
$u$.  Since $\Sigma-F^{-1}$ is real and symmetric, this implies that
$\Sigma-F^{-1}$ is positive-semidefinite; the matrix inequality
\begin{align} \label{MI}
\Sigma\ge F^{-1}
\end{align}
is the~\hbox{QCRB} in its most general form.  To use a CRB in
practice, it is customary to define a non-negative, quadratic cost
function $C \equiv\trace(\Lambda^T\Sigma)$ using a
positive-semidefinite (Hermitian) cost matrix $\Lambda$ suited to the
application \cite{vantrees,helstrom}.  The matrix QCRB is equivalent
to a lower bound, $C\ge\trace(\Lambda^TF^{-1})$, on all such cost
functions.

To calculate the QCRB, we must be more specific about the evolution
of the enlarged quantum system.  The Hamiltonian that governs overall
system dynamics over the interval $t_j\le t\le t_{j+1}$, of duration
$\delta t$, is $H_j(x_j)$, with corresponding evolution operator
$U_j=\exp[-iH_j(x_j)\delta t/\hbar]$.  We have $\partial U_j/\partial
x_j=U_j(-ih_j\,\delta t/\hbar)$, where $h_j\equiv
\partial H_j/\partial x_j$.  Let $U_{kj}\equiv U_{k-1}\cdots U_j$
denote the evolution operator over the interval $t_j\le t\le t_k$.
The density operator $\rho_x$ is related to the initial density
operator $\rho_0$ by $\rho_x=U_{J0}\rho_0 U^\dagger_{J0}$, which
gives $\partial\rho_x/\partial x_k=-i[M_k,\rho_x]$, where
\begin{align}
M_k\equiv i\parti{U_{J0}}{x_k}U^\dagger_{J0}=
\frac{\delta t}{\hbar}U_{Jk}h_kU^\dagger_{Jk}=
\frac{\delta t}{\hbar}U_{J0}\hat h_kU^\dagger_{J0}\;,
\label{Mk}
\end{align}
with $\hat h_k\equiv U_{k0}^\dagger h_kU_{k0}=h(t_k)$ being the
Heisenberg-picture version of $h_k$.  An obvious choice for $Q_k$ is
the anti-Hermitian $Q_k=-2i\Delta M_k$, where $\Delta M_k\equiv
M_k-\langle M_k\rangle_x$.
The quantum part of the Fisher matrix then becomes
\begin{align}
F_{jk}^{(Q)}\!=\!
\frac{4(\delta t)^2}{\hbar^2}\int\!\!Dx\,P[x]
\frac{1}{2}
\trace\!\Big((\Delta\hat h_j\Delta\hat h_k+\Delta\hat h_k\Delta\hat h_j)\rho_0\Bigr)\;,
\label{FQ2}
\end{align}
where $\Delta\hat h_k\equiv\hat h_k-\langle\hat h_k\rangle_0$.  Angle
brackets with subscript 0 denote an expectation value with respect to
$\rho_0$.  The quantum Fisher information is thus a two-time
covariance function, averaged over $P[x]$.

To take the continuous-time limit, we let $\delta t\to 0$,
$\Sigma_{jk}\to \Sigma(t_j,t_k)$, $F_{jk}/(\delta t)^2\to
F(t_j,t_k)$, and $\Lambda_{jk}/(\delta t)^2\to \Lambda(t_j,t_k)$.
The estimation-error covariance matrix becomes the two-time
covariance function of estimation error, $\Sigma(t,t')$, and the
Fisher matrix becomes $F(t,t')=F^{(Q)}(t,t')+F^{(C)}(t,t')$, with
\begin{align}
F^{(Q)}(t,t')&=
\frac{4}{\hbar^2}\int\!Dx\,P[x]\nonumber\\
&\phantom{\int\!}\times\frac{1}{2}
\Bigl\langle\Delta h(t)\Delta h(t')+\Delta h(t')\Delta h(t)\Bigr\rangle_{\!0}\;,
\\
F^{(C)}(t,t')&=\int\!Dx\,P[x]
\frac{\delta\ln P[x]}{\delta x(t)}\frac{\delta\ln P[x]}{\delta x(t')}\;,
\end{align}
$\delta/\delta x(t)$ being the functional derivative.

In the continuous-time limit, the matrix QCRB retains the same form
as Eq.~(\ref{MI}), where  the continuous-time inverse is defined by
$\int_{t_0}^{t_J} dt'' F(t,t'')F^{-1}(t'',t') = \delta(t-t')$.  The
bound on a cost function becomes
\begin{align}
C \equiv \int dt\,dt'\,\Lambda(t,t')
\Sigma(t,t') \ge \int dt\,dt'\,\Lambda(t,t')F^{-1}(t,t')\;.
\label{cost}
\end{align}
Equation~(\ref{cost}), valid for any cost function, is the most
serviceable expression of our chief result. An important special case
is the point estimation error,
\begin{align}
\Pi(t)\equiv\Sigma(t,t)=
\langle[\tilde x(t)-x(t)]^2\rangle\ge F^{-1}(t,t)\;,
\end{align}
where angle brackets without a subscript denote an overall
quantum-classical average.


To illustrate the use of our QCRB, we consider the estimation of a
force $x(t)$ on a quantum harmonic oscillator. The Hamiltonian is $H
= p^2/2m+m\omega_m^2 q^2/2 -q x(t)$, with $q$ being the position
operator, $p$ the momentum operator, $m$ the mass, and $\omega_m$ the
resonant frequency. In this situation, we have $h(t)=\partial
H(x(t))/\partial x(t) = -q$, which leads to a quantum component of
the Fisher information,
\begin{align}
F^{(Q)}(t,t')=\frac{4}{\hbar^2}\frac{1}{2}
\Bigl\langle\Delta q(t) \Delta q(t')+\Delta q(t')\Delta q(t)\Bigr\rangle_{\!0}\;.
\label{FQ3}
\end{align}
The further average over $P[x]$ in Eq.~(\ref{FQ2}) can be omitted in
Eq.~(\ref{FQ3}) because $x(t)$ appears linearly in $q(t)$ and thus
drops out of $\Delta q(t)$.  If we assume that $x(t)$ is a Gaussian
process, the classical, prior-information component of the Fisher
information is the inverse of the prior two-time covariance function
of $\Delta x(t)$~\cite{vantrees}.

We now assume that all noise processes are stationary.  For a
stationary, zero-mean process $f(t)$, the covariance function $\langle
f(t)f(t')\rangle$ depends only on the time difference $\tau=t'-t$ and
can be Fourier-transformed to give the power spectral density
$S_f(\omega)\equiv\int_{-\infty}^\infty d\tau\langle
f(t)f(t+\tau)\rangle e^{i\omega\tau}$.  The choice
$\Lambda(t,t')=\exp[i\omega(t'-t)]/(t_J-t_0)$, together with taking
$t_0\to-\infty$ and $t_J \to \infty$, makes $C(\omega)$ the power
spectral density of the estimation error. The QCRB~(\ref{cost}) then
becomes a \emph{spectral uncertainty principle}:
\begin{align}
C(\omega)
\!\left(S_{\Delta q}(\omega)+\frac{\hbar^2}{4S_{\Delta x}(\omega)}\right)
\ge\frac{\hbar^2}{4}\;.
\label{QCRBspectral}
\end{align}
In the time-stationary case, the matrix QCRB is equivalent to
satisfying this spectral uncertainty principle for all $\omega$. A
bound on the point estimation error now follows from
$\Pi=\int_{-\infty}^\infty(d\omega/2\pi)C(\omega)$.

To proceed in our approach, we must specify the measurements that
extract the force information from the oscillator and include the
associated backaction.  Thus we now suppose that one performs
continuous position measurements, using, for example, a continuous
optical probe.  The observation process is $y=q+\eta$, and the
oscillator equations of motion are $dq/dt=p/m$ and $dp/dt=-m\omega_m^2
q + x + \xi$, where $\xi$ is the backaction noise. Here $\xi(t)$ and
$\eta(t)$ are like the quadrature components of an optical field,
obeying the canonical commutation relation
$[\xi(t),\eta(t')]=i\hbar\delta(t-t')$.  We assume $\xi$ and $\eta$
have zero mean; their spectra satisfy an uncertainty principle,
$S_\xi(\omega)S_{\eta}(\omega)\ge\hbar^2/4$~\cite{wiseman,braginsky}.

If we introduce a small amount of damping, $\Delta q(t)$ is the
inhomogeneous solution for $q(t)$, driven just by $\xi(t)$, and
becomes stationary.  In the limit of negligible damping, the spectrum
of $\Delta q$ becomes $S_{\Delta
q}(\omega)=|G(\omega)|^2S_\xi(\omega)$, where $G(\omega)\equiv
1/m(\omega_m^2-\omega^2)$ is the oscillator transfer function.  The
spectral uncertainty principle~(\ref{QCRBspectral}) now takes the
form
\begin{align}
C(\omega)
\!\left(|G(\omega)|^2S_\xi(\omega)+\frac{\hbar^2}{4S_{\Delta x}(\omega)}\right)
\ge\frac{\hbar^2}{4}\;.
\label{QCRBspectraltwo}
\end{align}
The corresponding bound on point estimation error is
\begin{align}
\Pi \ge\intall \frac{d\omega}{2\pi}
\bk{\frac{4}{\hbar^2}|G(\omega)|^2S_\xi(\omega)+
\frac{1}{S_{\Delta x}(\omega)}}^{-1}\;.
\label{QCRBoscillator}
\end{align}
Notice that a bandwidth constraint on $x(t)$ is incorporated in the
prior information: $S_{\Delta x}(\omega)$ goes to zero outside the
relevant bandwidth, thus allowing $C(\omega)$ to be zero there and
making the integral~(\ref{QCRBoscillator}) finite.

We can elucidate the meaning of the QCRB~(\ref{QCRBoscillator}) by
considering how to estimate the force from the observations in this
scenario.  In the frequency domain, the observation process $y(t)$
reads $y(\omega) = G(\omega)[x(\omega) + z(\omega)]$, $z$ being a
noise term that depends on $\xi$ and $\eta$.  Using
smoothing~\cite{smooth,vantrees} to estimate $x$ from $y$ yields an
error
\begin{align}
\Pi &= \intall \frac{d\omega}{2\pi}
\bk{\frac{1}{S_z(\omega)}+\frac{1}{S_{\Delta x}(\omega)}}^{-1}\;.
\label{smooth_error}
\end{align}
This is the minimum achievable error for a given noise spectrum
$S_z(\omega)$. It cannot be reached by the more well-known technique
of filtering~\cite{wiseman}, as filtering does not make use of the
entire observation record.  If $\xi$ and $\eta$ are uncorrelated and
quantum limited, we have
\begin{align}
S_{z}(\omega)=\frac{S_\eta(\omega)}{|G(\omega)|^2}+S_\xi(\omega)
\ge\frac{\hbar}{|G(\omega)|}
\equiv S_{\rm SQL}(\omega)\;,
\label{SQL}
\end{align}
where the power spectrum $S_{\rm SQL}(\omega)$ is known as the
\emph{standard quantum limit\/} (SQL) for force
detection~\cite{braginsky}.

It is now evident that to attain the QCRB~(\ref{QCRBoscillator}), it
is necessary to beat the \hbox{SQL}. This requires evading or
tempering the effects of the backaction $\xi$.  One way to do this
is to correlate $\xi$ and $\eta$, as was proposed for interferometric
gravitational-wave detectors by Unruh~\cite{unruh}. An alternative is
to use quantum noise cancellation (QNC)~\cite{qnc}, which has the
advantage of making the QCRB~(\ref{QCRBoscillator}) achievable, as we
now show.  One QNC approach, discussed in~\cite{qnc}, adds an
auxiliary oscillator with position $q'$ and momentum~$p'$. One
monitors continuously the collective position $Q=q+q'$, giving a
process observable $y=Q+\eta$; the backaction force $\xi$ acts on
$P=(p+p')/2$ and thus equally, with strength $\xi$, on each of the
two oscillators.  Suppose the auxiliary oscillator has the same
resonant frequency and equal, but opposite mass (the negative mass
can be simulated by an optical mode at the red sideband of the
optical probe).  The dynamics of the collective position is then
determined by $dQ/dt=\delta p/m$ and $d\delta p/dt=-m\omega_m^2 Q+x$,
where $\delta p=p-p'$. There being no backaction noise in $z(t)$,
one easily finds that
\begin{align}
S_z(\omega)=\frac{S_\eta(\omega)}{|G(\omega)|^2}
\ge\frac{\hbar^2}{4S_\xi(\omega)}\frac{1}{|G(\omega)|^2}\;,
\end{align}
with equality for quantum-limited noise.  This
quantum-noise-cancellation scheme beats the SQL and if the noise is
quantum limited, does so optimally: the smoothing error given by
Eq.~(\ref{smooth_error}) achieves the QCRB~(\ref{QCRBoscillator}),
which implies that the spectral uncertainty
principle~(\ref{QCRBspectraltwo}) is saturated.  Our force-sensing
QCRB, rigorously proven and demonstrably achievable, thus serves as a
fundamental quantum limit, against which the optimality of future
force sensing schemes should be tested.  More generally, our QCRB for
arbitrary cost functions~(\ref{cost}) will find application whenever
quantum-limited estimation of temporally varying waveforms is
attempted.

We acknowledge productive discussions with J.~Combes, A.~Tacla,
Z.~Jiang, S.~Pandey, M.~Lang, and J.~Anderson.  This work was
supported in part by NSF Grants No.~PHY-0903953 and No.~PHY-1005540,
ONR Grant No.~N00014-11-1-0082, and ARC Grant~CE0348250.

\end{document}